\documentclass[journal,12pt]{IEEEtran}
\usepackage{times}

\usepackage{enumitem}

\usepackage{cite}

\usepackage[caption=false]{subfig}
\usepackage{graphicx}

\ifCLASSINFOpdf
   \DeclareGraphicsExtensions{.pdf,.jpeg,.png}
\else
   \DeclareGraphicsExtensions{.eps}
\fi
\begin{document}
%
\title{Towards Disaggregating the SDN Control Plane}
%
%
%

\author{\IEEEauthorblockN{
Douglas Comer, and
Adib Rastegarnia}

\thanks{Douglas Comer and Adib Rastegarnia are with the Department of Computer Science, Purdue University, West Lafayette,
IN,47906, USA, e-mail: (comer@cs.purdue.edu, arastega@purdue.edu}




}


\maketitle

\begin{abstract}
Current SDN controllers have been designed based on a monolithic approach that integrates all of services and applications into one single, huge program.  The monolithic design of SDN controllers restricts programmers who build management applications to the specific programming interfaces and services that a given SDN controller provides, making application development dependent on the controller, and thereby restricting portability of management applications across controllers. Furthermore, the monolithic approach means an SDN controller must be recompiled whenever a change is made, and does not provide an easy way to add new functionality or scale to handle large networks. To overcome the weaknesses inherent in the monolithic approach, the next generation of SDN controllers must use a distributed, microservice architecture  that disaggregates the control plane by dividing the monolithic controller into a set of cooperative microservices. In this paper, we explain the steps that are required to migrate from a monolithic architecture to a microservice architecture, and consider two potential designs that achieve the goal.  Finally, the paper reports the results of testbed measurements that we use to evaluate the proposed disaggregated architecture from  multiple  perspectives, including functionality and  performance.
\end{abstract}


%
\IEEEpeerreviewmaketitle

\section{Introduction}

Software defined networking (SDN) is an emerging trend for the design of Internet management systems that decouples vertical integration of the control plane and data plane and provides flexibility that allows software to program the data plane hardware directly according to a set of network policies. Thus, the control functions used to configure a device no longer need to be integrated with the functions that perform data forwarding. SDN offers flexibility to program and monitor computer networks directly using a \emph{controller} that runs software known as a \emph{Network Operating System (NOS)} to provide control logic. 
In the current SDN paradigm, management functionality crosses three key layers, including data plane forwarding mechanisms, control plane functions, and management applications \cite{7452335}. A centralized control plane is implemented by an \emph{SDN controller}. A controller uses two types of \emph{Application Program Interfaces (APIs)} to connect to other entities: a \emph{Northbound (NB) API} and a \emph{Southbound (SB) API}. A NB API defines communication between external management applications and the SDN controller; a SB API defines communication between the controller and underlying network devices \cite{7452335}. One of the most widely used SB APIs, known as the \emph{OpenFlow}  \cite{McKeown:2008:OEI:1355734.1355746} protocol, allows a controller to insert, modify, and update flow table rules, and to specify associated actions to be performed for each of the flows that pass through a given network device.  The architecture of current software defined management systems exhibits several weaknesses as follows:

\begin{itemize}[noitemsep,wide=0pt, leftmargin=\dimexpr\labelwidth + 2\labelsep\relax]

\item \textbf{Monolithic and Proprietary}: Current SDN controllers employ a monolithic architecture that aggregates all control plane subsystems into a single, gigantic, monolithic program. In the aggregated control plane model, each controller defines its own set of programming interfaces and services; when writing management applications, a programmer can only use the facilities the controller provides. An aggregated controller design makes application development dependent on a particular SDN controller and the specific programming language that has been used to implement the controller. Consequently it restricts portability across multiple controllers by making each application depend on a specific controller. In the monolithic approach, the vendor that creates a controller chooses all functionality and ensures the cohesion of all pieces. However, the approach does not provide an easy way for users to incorporate new services or adapt the controller quickly. 

\item \textbf{Lack of a Uniform Set of NB APIs}: 
 Even if they use the same general form of interaction (e.g.,  RESTful interface), SDN controllers, such as ONOS\cite{Berde:2014:OTO:2620728.2620744} and OpenDayLight\cite{odl}, each offer a NB APIs that differs from the APIs offered by other controllers in terms of syntax, naming conventions, and resources. The lack of uniformity among NB APIs makes each external management application dependent on the NB API of a specific SDN controller.

\item \textbf{Lack of Reusability of Software Modules}: In the current SDN architecture, the dependency between a management application and a specific type of controller limits reuse of SDN software. In many cases, when porting an application from one controller to another, a programmer must completely recode all modules, including basic modules that collect topology information, generate and install flow rules, monitor topology changes, and collect flow rule statistics. 

\item \textbf{Lack of Scalability and Reliability}: In current SDN controllers, there are strong dependencies between the subsystems. If one of the subsystems fails, the failure affects the operation of other susbsystems. In addition, a monolithic SDN controller is a gigantic and complex software that is difficult to scale because a given subsystem cannot be changed or scaled independently.

\end{itemize}
To overcome the above weaknesses, we propose a new architecture for the design and implementation of next generation SDN control plane systems that splits the current monolithic controller software into a set of cooperating  microservices. This article makes the following contributions:
\begin{itemize}
\item Explains the concept of \emph{SDN Control Plane Disaggregation} and introduce a distributed architecture for next generation of SDN controllers. 
\item  Describes steps taken towards disaggregating the SDN control plane, considers potential ways to achieve the goal, and discuss the advantages and disadvantages of each.  

\item Evaluates the approaches and consequent tradeoffs from multiple perspectives, such as performance and implementation effort.
\end{itemize}
The rest of the paper is organized as follows:   The next section summarizes related work. Section \ref{disaggregation} explains the concept of SDN control plane disaggregation. Section \ref{architectures} considers two event distribution systems that can potentially be used to externalize event processing. and Section \ref{grpcvskafka} compares the two event distribution systems by assessing implementation tradeoffs.
Section \ref{results} explains an experimental setup used to assess performance, and reports the results of measurements. Finally,  Section \ref{conclusion} concludes the paper.

\section{Related Work}
\label{relatedwork}
 Umbrella \cite{Comer:2018:UUS:3230718.3233546} is a unified software defined network programming framework  that provides a new set of APIs for  the implementation of SDN applications independent of the NB APIs used by specific SDN controllers. Umbrella uses OFtee \cite{oftee} as a tool to provide OpenFlow \emph{PACKET\_IN} messages to the external applications to support external reactive SDN applications as well as proactive applications.
 This article is an extension of \cite{epp}, which presents the idea of externalization of packet processing in SDN, one of the first steps towards control plane disaggregation.
This article generalizes the idea of external packet processing to external event processing, a design that allows complete disaggeration of control plane services. In addition, this article focuses on required steps to migrate from a monolithic SDN control plane architecture to a disaggregated control plane.

\section{SDN Control Plane Disaggregation}
\label{disaggregation}
\begin{figure*}[!ht]
\centering
\includegraphics[scale=0.62]{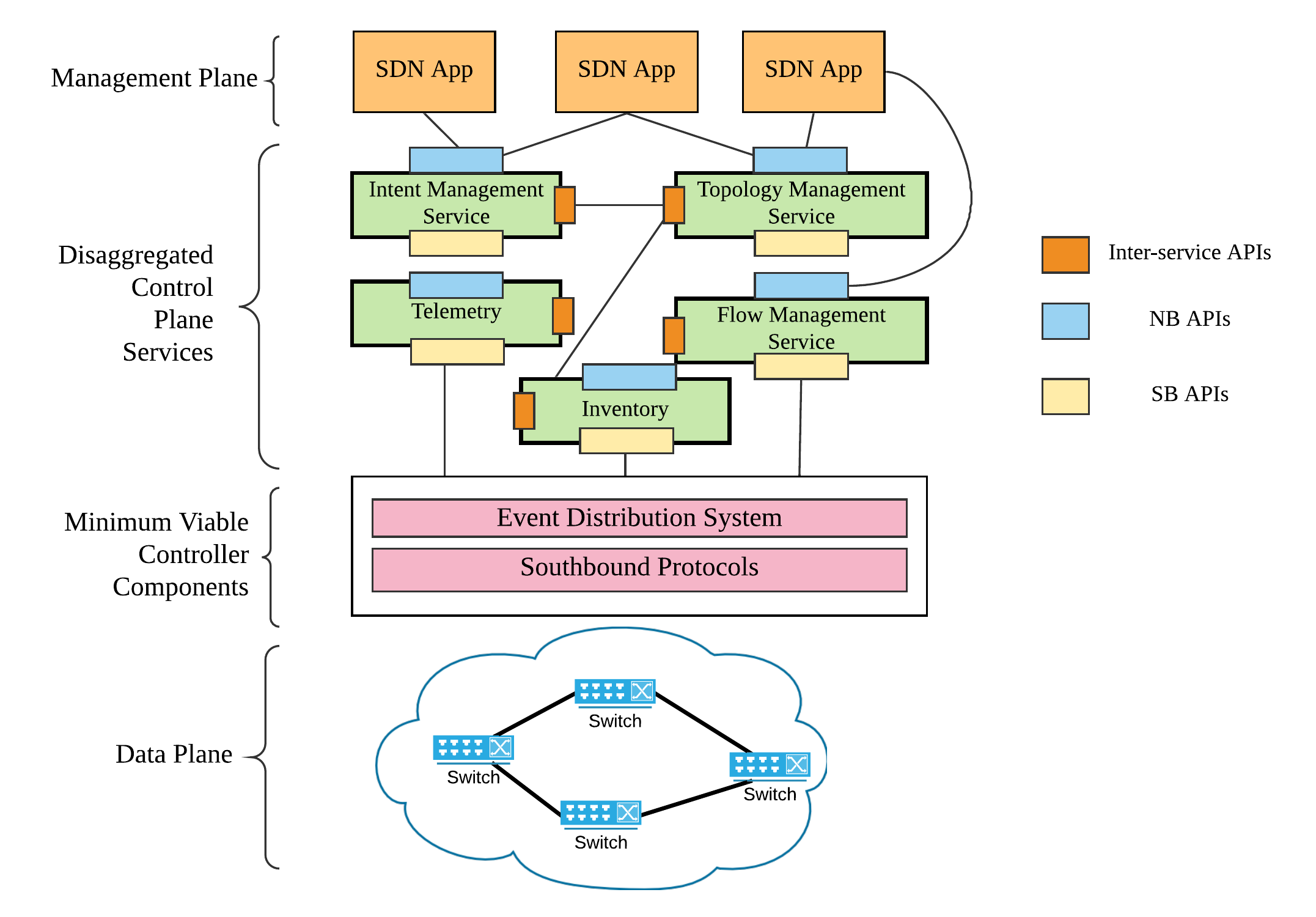}
\caption{The Disaggregated SDN Control Plane Architecture}
\label{disaggregated_ctrl_plane}
\end{figure*}

As Figure \ref{disaggregated_ctrl_plane} illustrates, a monolithic SDN controller can be disaggregated into a suite of cooperating \emph{microservices}, such that a controller \emph{core} provides the minimum required functionality, and microservices, that exist outside of the controller core provide all other services.  Migrating from a monolithic controller to a microservice architecture for control planes offers the following benefits: 

\begin{itemize}
    \item \textbf{Flexibility In Scaling}: One of the main advantages of the microservice architecture arises from its ability to scale a given service horizontally, independent of other subsystems and services. 
    \item \textbf{Freedom In Choosing a Programming Language}:  In the disaggregated control plane model, programmers have the opportunity to choose an arbitrary programming language, programming technology,  and third-party library function when building a given SDN management application.  The approach makes the application development process more flexible, and allows a programmer to choose a language that is appropriate to a given app. 
    \item \textbf{Fault Isolation}: In current SDN controllers, if one of the subsystems fails it can affect the entire controller. A disaggregated control plane means that the failure of a given microservice will not affect other microservices.  Moreover, a microservice can be repaired and restarted without recompiling the controller and without restarting other microservices.
    \item \textbf{A Controller Core With Minimal Components}: In the monolithic approach, every instance of a controller includes all services and apps, even if the instance only needs a small subset. In the disaggregated model, the controller core contains the minimum viable set of components and functions, and only the required apps and services need to be deployed outside of the controller core.
    \item \textbf{A Disaggregated Code Base}: In the monolithic architecture, a single, large code base includes all services and apps. Consequently, changes to even a small, seldom-used service requires changing the controller code base. In the disaggregated model, each services and each applications can be in a separate code base, isolating changes.
\end{itemize}

The key components of a disaggregated control plane architecture consist of:
\begin{itemize}
    \item \textbf{Minimum Viable Controller Components}: The first step towards SDN control plane disaggregation involves identifying a minimal set of viable controller core components. When it receives a request from a management app or from a control plane service, a control plane core service uses a SB protocol to communicate with the underlying network devices, and then uses the information to respond to the request. Consequently, a controller with minimum functionality needs the following two major subsystems: 
\begin{itemize}
   \item \textbf{An Event Distribution System}: This subsystem notifies microservices (i.e. control plane services and management applications) about events that occur in the underlying network devices, such as flow rule. network link, device, or packet exception events. The event distribution system provides a standard API used to stream events to the external processes (i.e. microservices); to permit multilingual apps, the API must be independent of the programming language used to implement each microservice.
    \item \textbf{SB Protocols}: SB protocols define the communication between underlying network devices and the core of the controller. The event distribution system communicates uses SB protocols to receive events from underlying devices, and then uses the distribution mechanism to propagate event notifications to external microservices.

\end{itemize}
    \item \textbf{Disaggregated Control Plane Services}: In the disaggregated model, each of the core control plane services (e.g. topology service, flow service, etc) is a microservice that can run in a container.  The microservices (i.e. the containers in which the service runs) can be orchestrated using a conventional container orchestrator technology, such as \emph{Kubernetes} \cite{kubernetes}. Each control plane service provides three sets of APIs: NB APIs that define the communication between control plane services and applications, inter-service APIs that define the communication between each microservice,  and SB APIs that define the communication between each microservice and controller core components. 
    \item \textbf{Management Plane}: Management plane comprises set of management applications that use disaggregated control plane services to implement network control functions, policies, and strategies, such as routing, network monitoring, firewall rules, network address translation, and load balancing. To access microservice, management applications use the NB APIs that the microservices offer. In fact, each management application forms a microservice that runs in a container exactly like other microservices.
\end{itemize}

\section{An Overview of Event Distribution Mechanisms}
\label{architectures}
We define an event as a change in the network. For example, consider the following network events that occur commonly: 
\begin{itemize}
    \item \textbf{Packet Exception Event}. A switch is configured to send any packet to the controller core if the packet does not match any of the installed forwarding rules.  A \emph{packet exception event} occurs when a packet arrives at the controller core.
    \item \textbf{Topology Event}. Whenever the network topology changes (e.g., a link fails or a link is placed back in service), the controller core receives a \emph{topology event}. Topology events can be categorized according to type: link, device, port, and so on. 
    \item \textbf{Flow Rule Event}: Flow rule event occurs whenever the flow rules change (e.g, when a rules if added, removed, or updated).
\end{itemize}
As the previous section explains, disaggregating control plane requires external event processing. A  typical  SDN  controller provides  several  built-in  services, such as a topology discovery service. Provided that event processing can be externalized, most of the control plane services in an SDN controller can be implemented as microservices outside of the controller core. For example, to implement a topology discovery microservice, the event distribution mechanism can capture and forward both \emph{link layer discovery protocol (LLDP)} and \emph{Address Resolution Protocol (ARP)} packets to an external microservice that uses the packets to build a map of links between switches and end-hosts.
Both a topology service and management apps can learn about changes in connectivity from link and device events. Consequently, the topology microservice and management apps can each react to the network changes. The next subsections consider the implementation of event distribution mechanisms using \emph{publish-subscribe} and \emph{point-to-point} messaging paradigms.

\subsection{An Event Distribution Mechanism Using A Publish-Subscribe Model}
\begin{figure}[ht]
    \centering
    \includegraphics[width=\columnwidth]{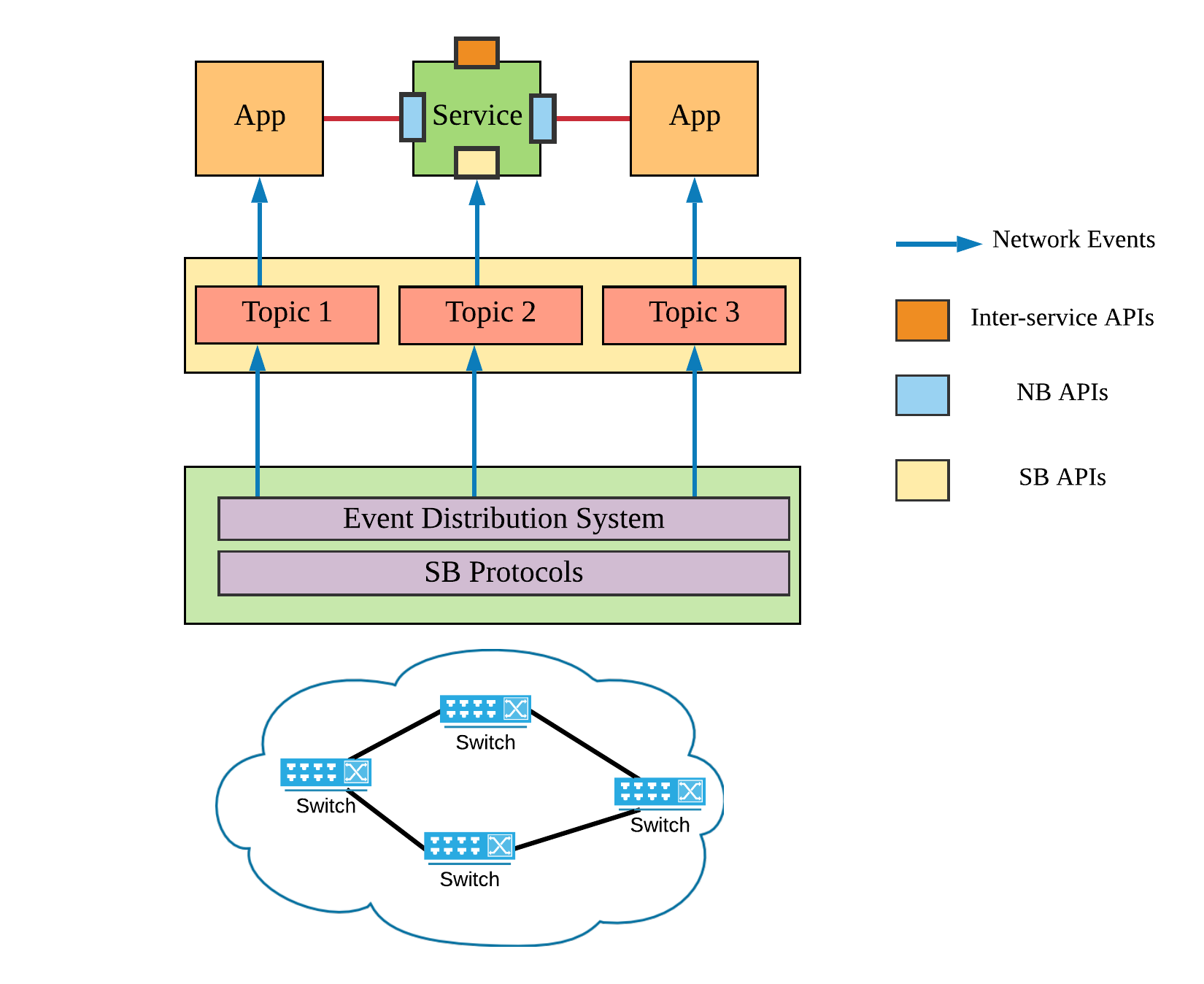}
    \caption{The architecture of the event distribution system using publish-subscribe paradigm}
    \label{fig:event_distribution_publish_subscribe}
\end{figure}

As Figure \ref{fig:event_distribution_publish_subscribe}  illustrates, the event distribution system that runs inside the controller listens for events that occur in the controller core, and pushes each events into a topic in an event broker where external processes and  applications can consume the event by  pulling it from a topic in the event broker.  In essence, the event distribution system acts as a producer by pushing items to  brokers,  and  external  applications  and  services  act  as  consumers that receive events from the broker. In the publish-subscribe model, each type of event is associated with a \emph{Topic}.  Thus, an external application or a service specifies a topic, and then receives events sent to the topic.  If an app wishes to receive packet events, the app subscribes to the packet event topic.  Once an app subscribes to a topic, the app will receive a notification each time the event distribution systems publishes an event on the topic.

\subsection{An Event Distribution Mechanism Using A Point-to-Point Model}
\begin{figure}[ht]
    \centering
    \includegraphics[width=\columnwidth]{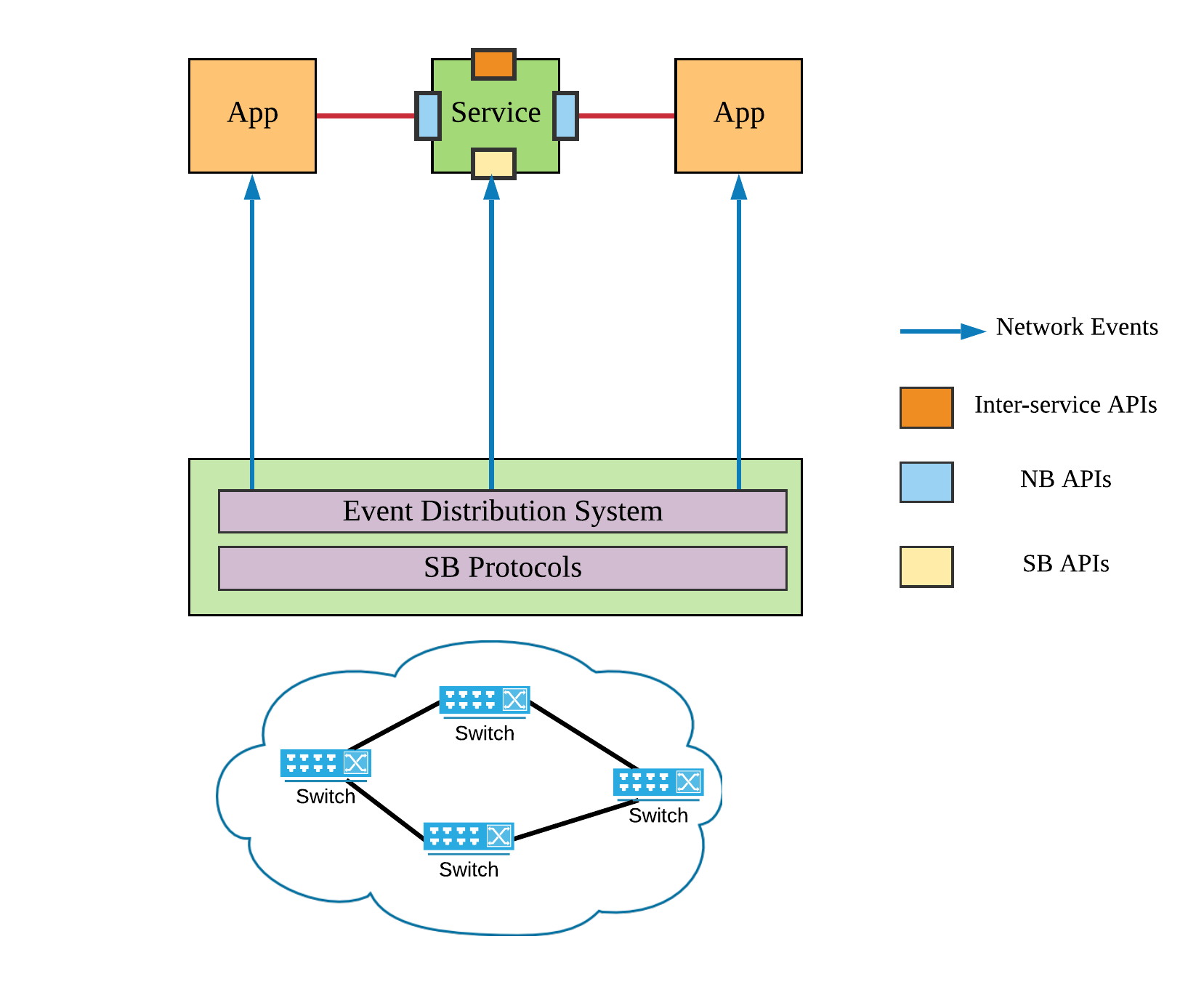}
    \caption{The architecture of the event distribution system using point-to-point paradigm}
    \label{fig:event_distribution_point_to_point}
\end{figure}

As Figure \ref{fig:event_distribution_point_to_point} illustrates, an event distribution system that follows a point-to-point model uses a point-to-point channel to send event notifications to each of the applications and services. The event distribution system employs a push-notification mechanism that provides a copy of each incoming events to each external services or app that has subscribed to the type of the event.  In essence, the the event distribution system streams events to each of the subscribers.

\section{Implementation Options and Trade-offs}
\label{grpcvskafka}
Various technologies and frameworks that can be used to implement microservice architectures, including container technologies such as Docker, container orchestrators such as Kubernetes, and microservice frameworks. In a microservice architecture, microservices communicate with each other via standard APIs such as gRPC or a REST API.  In addition, one can use frameworks such as Apache Kafka, RabbitMQ, and ActiveMQ to implement an event distribution system that follows the publish-subscribe model; frameworks, such as gRPC, can be used to implement an event distribution system that follows the point-to-point model.
Studying and comparing of all potential implementation technologies falls outside the scope of this paper. However, to investigate the feasibility of the proposed disaggregated control plane architecture,
we report some preliminary results based on Kafka \cite{kafka} and gRPC \cite{grpc}, two well-known frameworks.
In both implementations, we use protocol buffers \cite{protobuf} that are a language-neutral, platform-neutral extensible mechanism for serializing structured data to encode network events. For the purpose of experimental results, we use a gRPC API to return the packet to the controller core, which will send the packet to a switch. Furthermore, If an application or service needs to install flow rules on network devices it has a choice of using a REST API or gRPC. 

\subsection{A Comparison Of Kafka and gRPC Event Distribution Mechanisms}

The advantages and disadvantages of the two event distribution mechanisms can be summarized: 

\begin{itemize}
    \item \emph{Programming Language Agnosticism}: Each of the proposed event distribution mechanisms provide a facility that can be used to implement services and applications in arbitrary programming languages. From the client side perspective, however, gRPC makes it easier than Kafka to expand a distribution system to include apps written in new programming languages. To use gRPC with a new language, a programmer only needs to compile the set of protobuf messages used for communication between the event distribution system and each microservice. The gRPC technology automatically generates the gRPC stubs that external SDN applications and services need. In contrast, to use Kafka with a new programming language, a programmer must implement a set of high level abstractions that applications use to receive Kafka events. 

    \item \emph{Loose Coupling}: Kafka, and the publish-subscribe paradigm in general, provides a more loosely coupled communication system than gRPC and point-to-point messaging technologies. In  Kafka's publish-subscribe paradigm, producers and consumers do not need to know the existence of other producers and consumers. In gRPC, however, a client needs to know about the services that the event distribution system provides. Using a loosely coupled event distribution system offers the advantages of scalability, resilience, and maintainability.
    
    \item \emph{Fault Isolation and Code Base Disaggregation}:  Disaggregating the control plane services and using an event distribution mechanism increases fault isolation and allows code base disaggregation regardless of the language used, by keeping components independent of one another.
   
\end{itemize}

\section{Experimental Results}
\label{results}
This section presents an evaluation of the Kafka and gRPC distribution mechanisms using various performance metrics, primarily response time and throughput. 

\subsection{Experimental Setup} 
We implemented an early of version of the proposed event distribution systems as an application for the ONOS \cite{Berde:2014:OTO:2620728.2620744} SDN controller. To measure the two distribution mechanisms, we used an SDN testbed that consists of 10  OpenFlow  switches  that  logically  define  5  interconnected sites.  We use the \emph{virtualized mode} feature on the network switches to divide each physical switch into 10  independent smaller switches.  Each emulated site implements a Fat-tree network topology.

\subsection{Experimental Scenarios}
    \subsubsection{ \textbf{Response Time}} This experiment measures response time, and provides an evaluation of the overhead of external event processing. Our goal is to compare the amount of time that an external app or service needs to process a packet event with the time it takes to process the same packet even inside the monolithic version of ONOS, and to understand the effect on overall response time.  In the internal ONOS packet processing, whenever a host sends a ping request for which no forwarding rules have been established, each switch along the  path  sends the incoming  packet to the  controller core, which  processes the  packet internally and returns the packet to the switch to be forwarded.  In the  disaggregated architecture, whenever a host sends a ping request for which no forwarding rules have been established, each switch along the path sends the incoming packet to the controller core, and the event distribution system in the core (using either Kafka or gRPC) distributes the incoming packet to the set of external apps and services that have subscribed to packet events. The external process then uses gRPC to return the packet to the switch for forwarding.  To focus measurements on the control plane overhead, we did not install flow rules.  Thus, each packet causes a packet event at each switch along the path.  We ran the experiment 500  times  and measured the ping response time between two end hosts that are 5 hops apart in our SDN testbed.  Externalization of packet processing introduces overhead that increases the overall response time. As a baseline, we measured the average response time for internal processing in ONOS as $24$ ms.
    The average response time for a gRPC system is $29$ ms, and the average time for a Kafka system is $35$ ms. We observe that using gRPC introduces less overhead than the Kafka event distribution mechanism. 
    \subsubsection{ \textbf{Throughput}} 
\begin{figure}
    \centering
    \includegraphics[width=\columnwidth, trim={0 0.1cm 0 0},clip]{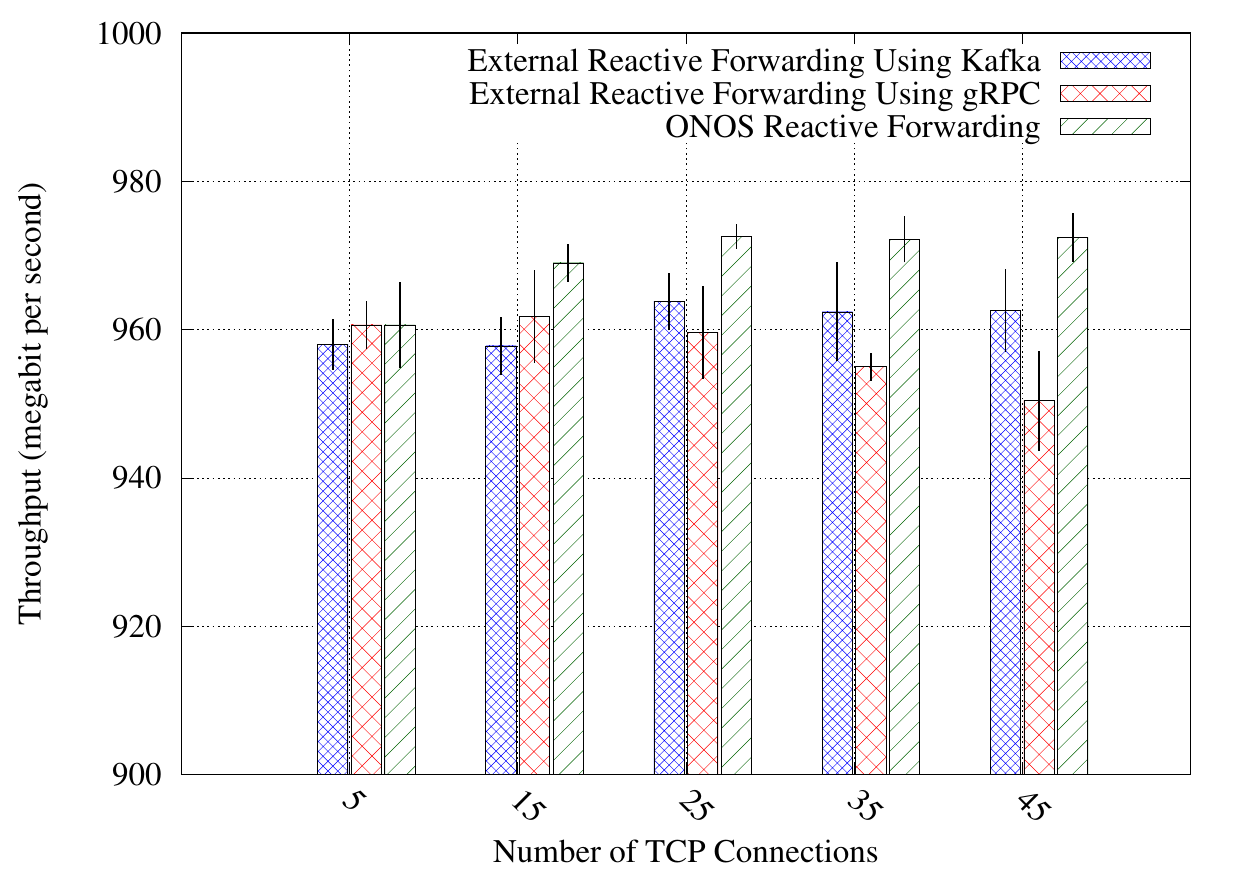}
    \caption{Throughput vs. number of TCP connections for external and internal ONOS reactive forwarding applications}
    \label{fig:throughtput}
\end{figure}

To assess the impact of  externalized packet processing and the use of a REST API for flow rule installation on throughput, we compare external reactive forwarding applications that use gRPC and Kafka with the same reactive forwarding application compiled into ONOS. In all three cases,  we use  a  hard time-out  of  10  seconds  to  remove  all flow  rules  (i.e.,the flow rules installed in a switch disappear every 10 seconds.  Following removal of the flow rules, the next packet to arrive will causes a packet event, and the subsequent re-installation of the rules).  We  use  the  iperf3  tool  to generate TCP traffic, varying the the number of concurrent TCP connections, and running each measurement for 150 seconds. As the results in Figure \ref{fig:throughtput} show, the effect of externalized packet processing on throughput is negligible. Moreover, the overhead of using a REST API to install flow rules is larger than overhead introduced by externalization.

\section{Conclusion}
\label{conclusion}
A monolithic architecture for an SDN controller aggregates all control plane subsystems into a single, gigantic program. An SDN controller that adopts the monolithic approach restricts programmers who write management applications to the programming interfaces and services that the controller provides. To overcome the limitations inherent in a the monolithic architecture, we propose a distributed architecture that disaggregates controller software into a small controller core and a set of cooperative microservices. Disaggregation allows a programmer to choose a programming language that is appropriate for each microservice.  To migrate from a monolithic approach to a disaggregated microservice architecture, a mechanism must be devised to distribute events to external processes. In this paper, we evaluate two candidate distribution mechanisms: Kafka and gRPC. Our experimental results show that externalizing event processing introduces some overhead, and the overhead resulting from gRPC is lower than the overhead resulting from Kafka.  Externalizing event processing has a negligible effect on throughput, and the cost is considered small when compared with the advantages of portability, flexibility, and support for multilanguage management applications.

\ifCLASSOPTIONcaptionsoff
  \newpage
\fi


\bibliographystyle{IEEEtran}
\bibliography{ref.bib}
\vskip -2.1\baselineskip plus -1fil
\begin{IEEEbiography}[{\includegraphics[width=1in,height=1.2in,clip,keepaspectratio]{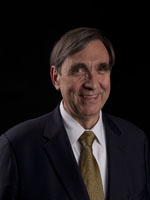}}]{Douglas Comer}
is an internationally recognized expert on computer networking and the TCP/IP protocols. He has been working with TCP/IP and the Internet since the late 1970s. Comer established his reputation as a principal investigator on several early Internet research projects. He served as chairman of the CSNET technical committee, chairman of the DARPA Distributed Systems Architecture Board, and was a member of the Internet Activities Board (the group of researchers who built the Internet). Professor Comer is well-known for his series of ground breaking textbooks on computer networks, the Internet, computer operating systems, and computer architecture. His books have been translated into sixteen languages, and are widely used in both industry and academia.
\end{IEEEbiography}
\vskip -2.1\baselineskip plus -1fil
\begin{IEEEbiography}[{\includegraphics[width=1in,height=1.2in,clip,keepaspectratio]{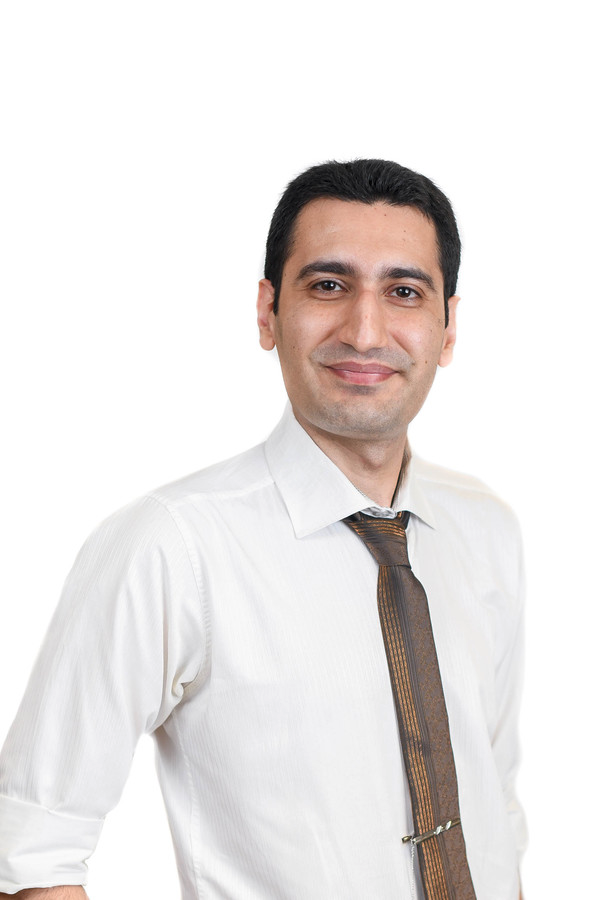}}]{Adib Rastegarnia}
is a PhD candidate at Computer Science Department of Purdue University under the supervision of Prof. Douglas Comer. He earned his Master’s degree in Computer Science from Purdue University in 2018. His current research interests span the areas of computer networks, operating systems, SDN, and Internet of Things (IoT) with a focus on designing and implementing working prototypes of large, complex software systems and conducting real world measurements. He is also a member of Systems Research Group of Computer Science Department at Purdue University. 
\end{IEEEbiography}

%




\end{document}